# Responsibility Modelling for the Sociotechnical Risk Analysis of Coalitions of Systems

David Greenwood and Ian Sommerville
Dependable Systems Engineering Group
University of St Andrews
St Andrews, Scotland

*Abstract*—Society is challenging systems engineers by demanding ever more complex and integrated systems. With the rise of cloud computing and systems-of-systems (including cyber-physical systems) we are entering an era where mission critical services and applications will be dependent upon 'coalitions-of-systems'. Coalitions-of-systems (CoS) are a class of system similar to systems-of-systems but they differ in that they interact to further overlapping self-interests rather than an overarching mission. Assessing the sociotechnical risks associated with CoS is an open research question of societal importance as existing risk analysis techniques typically focus on the technical aspects of systems and ignore risks associated with coalition partners reneging on responsibilities or leaving the coalition.

We demonstrate that a responsibility modeling based risk analysis approach enables the identification of sociotechnical risks associated with CoS. The approach identifies hazards and associated risks that may arise when relying upon a coalition of human/organizational/technical agents to provision a service or application. Through a case study of a proposed cloud IT infrastructure migration we show how the technique identifies vulnerabilities that may arise because of human, organizational or technical agents failing to discharge responsibilities.

## I. INTRODUCTION

Society is challenging systems engineers by demanding ever more complex and integrated systems. With the rise of cloud computing and systems-of-systems (including cyber-physical systems) we are entering an era where mission critical services and applications will be dependent upon 'coalitions-of-systems' [1, 2]. Coalitions-of-systems (CoS) are a class of system, similar to systems-of-systems, that comprise a set of systems that interact to further overlapping self-interests rather than an overarching mission [3, 4]. A distinctive trait of CoS is that their continuing operation is dependent upon the potentially fragile overlap of the coalition's self-interests and therefore proactive risk management is essential [5].

If future mission critical services or applications are to be deployed as CoS then techniques are required to identify and analyze the risks of making coalition members responsible for provisioning parts of services or applications and the liabilities that may be incurred. This challenge is of academic importance as it raises many questions with respect to the development of abstractions that are practical and scalable enough to analyze large distributed sociotechnical systems [6-8].

In the near future a typical CoS scenario may comprise an enterprise using a commercial public cloud IT infrastructure to provision a service for customers. The parts of this CoS comprise the cloud provider's infrastructure, the enterprise's IT systems to monitor and manage virtual machines on the cloud infrastructure, the ISPs providing connectivity, and the customer's IT systems consuming the service (perhaps to monitor and manipulate a physical process). This system meets the criteria of a CoS as it is held together by the following overlapping self-interests: to generate profit (cloud provider, enterprise, ISP); to obtain a desired service at a competitive price (customer). Like all CoS this system is fragile as the situation is dependent on the situation not changing in ways that a make the coalition's self-interests non-overlapping e.g. significant changes in price or service quality.

We propose that modeling the responsibilities of the agents in a system is a promising avenue for analyzing the risks associated with CoS. This is because entering into a coalition necessarily involves the reliance on other parties to discharge responsibilities appropriately such that services and applications meet service level agreements or safety standards.

In this paper we demonstrate the use a responsibility modeling based risk identification approach to identify risks associated with the use of a commercial public cloud IT infrastructure by an enterprise to provision a service for customers in the Scottish North Sea oil & gas sector.

## II. COALITIONS OF SYSTEMS AND SOCIOTECHNICAL RISK

Systems-of-systems are a class of system whose interacting parts comprise systems, that are owned and managed by independent parties, and whose parts evolve over time [9]. Typical examples are integrated supply chain management systems, integrated healthcare networks, and cyber-physical systems such as integrated embedded systems within ships, land vehicles, aircraft, or industrial plants. Coalitions-of-systems (CoS) are a class of system similar to systems-of-systems but they differ in that they interact to further overlapping self-interests rather than an overarching mission.

Sociotechnical risks are an important factor when analyzing the risks associated with CoS as the self-interests that hold the system together may be fragile. For example, a CoS formed

We thank the EPSRC (grant numbers EP/H042644/1 and EP/F001096/1) for funding this work.

around the use of a commercial public cloud infrastructure may be sensitive to changes in pricing or service offerings. Therefore in addition to understanding the technical risks it is important to understand the sociotechnical dependencies between parties, the kinds of changes that could disrupt the coalition and the liabilities that may be incurred.

There are number of sociotechnical risk analysis approaches relevant to CoS. The most notable are Functional Resonance Analysis Method (FRAM) [10, 11], Systems-Theoretic Accident Modeling and Processes (STAMP) [12, 13] and Responsibility Modeling for Risk Analysis [14, 15].

The FRAM approach is a process level accident analysis and risk analysis method. It uses the concept of interacting 'functions' to represent a process and identify risks to the outcome of a process. FRAM is performed by divided up a process into a number of interacting functions comprising inputs, outputs, preconditions, control constraints, timing constraints and resources. The potential variability of functions is identified and the implication of this variability is determined by identifying its consequences on the outcome of the process. FRAM has been used to identify risks and analyze accidents involving sociotechnical systems including air accidents and medical accidents [16].

The STAMP approach is an institutional level accident analysis and risk analysis method. It uses the concept of interacting parts in dynamic equilibrium to represent institutional structures and identify risks in terms of 'control problems'; the premise being that risks arise when a system's behavior is not appropriately measured/detected and controlled. Identified risks are analyzed in terms of their interactions with the systems control structures and their resultant effect on institutional outcomes. STAMP has been used to analyze high profile sociotechnical accidents including the loss of Space Shuttle Columbia and the Walkerton water contamination tragedy [17].

The responsibility modeling approach is a tactical level risk analysis method. It uses the concept of responsible agents (human / organizational agents) and their interactions to represent a situation and identify risks in terms failures of agents to fulfill responsibilities. Risk identification is performed by: firstly modeling the responsibilities of the agents involved in a situation and the resources they require to discharge their responsibilities; secondly identifying the consequences/liabilities resulting from of an agent not having a resource or not discharging a responsibility. This is facilitated by the use of hazard keywords such as early, late, never, incapable, insufficient and impaired. Responsibility modeling has been used to analyze the failure of sociotechnical systems including E-counting systems in the Scottish elections and UK civil emergency planning [14, 15].

We believe that responsibility modeling is a technique that is complementary to both STAMP and FRAM. STAMP may be used to identify risks at an institutional governance level whilst FRAM may be used to identify risks at a process level. Responsibility modeling enables the analysis of situations at a 'tactical level' by understanding the risks related to agents depending upon others to discharge their responsibilities.

We believe that responsibility modeling offers a number of attractive characteristics that make it more suitable for sociotechnical risk analysis of CoS than either FRAM or STAMP. Firstly the responsibility abstraction provides a natural way of identifying the risks associated with CoS as entering into a coalition necessarily involves the reliance on other parties to discharge responsibilities. Secondly responsibilities are relatively unproblematic to elicit as people find them 'natural' to articulate in comparison to 'technical' constructs such as functions or goals. Thirdly responsibility modeling is relatively rapid to perform unlike FRAM. FRAM is concerned with process level risks and therefore elicits information such as functions, preconditions, control constraints and so on, which may be impractical for large-scale systems. Fourthly, STAMP is unsuitable as it focuses on institutional level control structures to identify 'control problems'. This assumes that there already exist appropriate techniques to identify/detect risks with CoS therefore STAMP is not a candidate technique.

Despite responsibility modeling having some similarities to goal-based risk analysis approaches, such as Astrolabe [18], it differs significantly. The concept of a responsibility embodies the notion that it is important *how* an agent acts. For example, a doctor that has performed procedures in accordance with legal and domain standards may have successfully discharged their responsibility for patient care even if their patient dies. Similarly if a patient lives but their treatment was unethical then the doctor will may be held liable. Unlike responsibility approaches, goal-based approaches principally focus on what has to be achieved, rather than foregrounding *obligations*, *liabilities* and conformance to *norms* or *standards*.

III. RESPONSIBILITY MODELING FOR RISK IDENTIFICATION

Responsibility modeling has been proposed by several researchers as a useful abstraction for analyzing the dependability of sociotechnical systems [19-21]. We use responsibilities as part of a graphical modeling notation that represents 'responsibilities' 'agents' and 'resources' interconnected by relationships.

For the purposes of responsibility modeling a responsibility is defined as:

"*A duty, held by some agent, to achieve, maintain or avoid some given state, subject to conformance with organizational, social and cultural norms*" [14]

The term duty in this definition captures obligation and accountability aspects of responsibilities such that if an agent does not appropriately discharge their obligation they will be held liable. The phrase conformance with organizational, social and cultural norms captures the fact that responsibilities must be discharged in accordance with legal and domain standards.

For the purposes of modeling the CoS analyzed in this paper we use the following entities and relationships:

**Responsibility**: An entity representing a duty to achieve, maintain or avoid a specified activity or state.

**Information Resource**: An entity representing a resource that provides information that contributes to meeting an obligation e.g. documents, databases.

**Physical Resource**: An entity representing a physical resource that contributes to meeting an obligation e.g. a server, tape drive, machine.

**Human Agent**: An entity representing a human being often referred to by their role e.g. Support Manager.

**Organizational Agent**: An entity representing an organization e.g. an enterprise or government agency.

**Responsibility For**: A relationship representing the allocation of a responsibility to an agent

**Has**: A relationship representing the allocation of a resource to an agent or responsibility

**Association**: A relationship representing that an entity is related to another. The association relationship may be annotated to clarify the relationship if necessary.

Responsibility modeling may be combined with a HAZOPS style approach to identifying risks [14]. Risks are identified via the means of 'risk clauses' that are composed from a target, hazard, condition and consequences.

**Target**: The entity or relationship to which the risk clause refers. E.g. An entity may be the responsibility to support and maintain a leased telecommunications line. A relationship may be the allocation of a responsibility or resource to an agent.

**Hazard**: Using a restricted set of keywords we aim to provide a checklist of hazard source categories to consider. The hazard keywords we used are outlined below:

- **Early** Occurrence of entity/relationship before required.
- **Late** Occurrence of entity/relationship after required.
- **Never** Non-occurrence of entity/relationship.
- **Incapable** Occurrence did not take place although attempts were made to fulfill the obligation.
- **Insufficient** Occurrence of the entity/relationship at an incorrect level.
- **Impaired** Occurrence of the entity/relationship in an incorrect manner.
- **Changes:** The entity/relationship changes on a permanent basis.

**Condition**: A description of the potential conditions that could manifest as a result of the hazard category considered.

**Consequences/Liabilities**: The potential effects or liabilities resulting from the hazard manifesting itself.

An example 'risk clause' may be found in Table 1 at the rear of this paper. In the context of analyzing CoS the entities and relationships under analysis are those between agents or resources from different organizations.

## IV. CASE STUDY

### A. The Situation

The case study organization is a UK based company (Company B) that provides bespoke IT solutions for the oil & gas industry. It comprises around 30 employees with offices in the UK and the Middle East. It has an organizational structure based on functional divisions (e.g. administration, engineering and support). We became involved with the organization as they were interested in exploring the cost saving opportunities that cloud computing could offer them. We therefore collaborated with the organization to assess the feasibility of the migration of one of the organization's primary service offerings (a quality monitoring and data acquisition system) to Amazon EC2 – an infrastructure-as-a-service offering from Amazon Web Services. Naturally, we were aware that the introduction of a new technology could be disruptive so we analyzed the sociotechnical risks of the proposed migration.

The situation was as follows: Company C is a small oil and gas company that owns some offshore assets in the North Sea oilfields. Company C needed a data acquisition system to allow them to manage their offshore operations by monitoring data from their assets on a minute-by-minute basis. Company C's assets rely on the production facilities of Company A (a major oil company), therefore the data comes onshore via Company A's communication links. Company C does not have the capabilities to develop their own IT systems; hence they outsourced the development and management of the system to Company B, which is an IT solutions company with a small data centre. Fig. 1 provides an overview of the system, which consists of two servers:

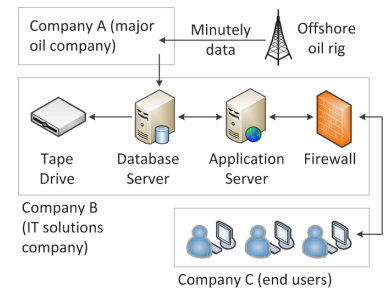

Figure 1. Logical structure of the 'as-is' system

1. A database server that logs and archives the data coming in from offshore into a database. A tape drive is used to take daily backups of the database, the tapes are stored off-site.

2. An application server that hosts a number of data reporting and monitoring applications. The end users at Company C access these applications using a remote desktop client over the Internet.

The system infrastructure was deployed in Company B's data centre and went live in 2005. Since then, Company B's support department have been maintaining the system and solving any problems that have risen. This case study investigated the risks of deploying the same system using the cloud offerings of Amazon Web Services. Fig. 2 provides an overview of this scenario, where Company B deploys and maintains the same system in the cloud.

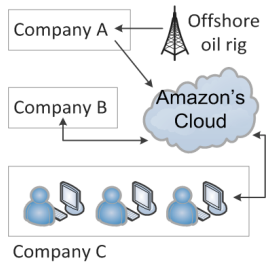

Figure 2. Logical structure of the 'to-be' system

*B. Fieldwork and Results*

We performed a series of interviews to identify key stakeholders and their responsibilities vis-à-vis the current 'as-is' situation and the proposed 'to-be' situation after migration to cloud. Our investigative remit was to consider the effect it would have on Company B. However, we did not have permission to speak to suppliers or customers thus limiting our interviewees to: a project manager; a technical manager; a support manager; two members of support staff; and a business development manager. The interviewees were encouraged to discuss their concerns regarding the proposed project and also the opportunities that it could afford. In order to organize and integrate the available information we created 'as-is' (Fig. 3) and 'to-be' (Fig. 4) responsibility models of the situation.

The purpose of these models is to provide a visual representation of the changes in stakeholder interdependencies and use this to structure the risk analysis. Agents are represented as triangular brackets. Responsibilities are represented as round edged rectangles. Resources in square brackets. The 'responsible for' relationship is represented by a line with a square end point. The 'has' relationship is represented by a line with a circular end point. As a form of shorthand entities (agents, responsibilities and resources) may be placed within a dashed container box and all those entities within the box share the relationship attached to the box. For example, in Fig. 3, the organizational agent 'Telco' is responsible for supporting the leased line and providing/maintaining their lease line pricing model.

Upon contrasting the 'as-is' (Fig. 3) situation with the 'to-be' (Fig. 4) situation we made a number of immediate observations.

- The introduction of EC2-style services would not only affect stakeholders involved in the provisioning of technical services but would have a direct effect on Finance/Business development, Sales/Marketing, and Customer relations staff.

- The support manager's responsibility of timely resolution of support calls would be ultimately dependent upon agents and resources no longer within his control (e.g. Amazon EC2 support services).

- In the 'as-is' situation, Company B has direct control of the in-house hardware on which the service executes and contractual relationship with the supplier of the leased line.

- In the 'to-be' situation, Company B is losing control over service quality, as it no longer has end-to-end control over the resources required for a customer to consume their service. The service would execute on Amazon hardware and customers (Company C) would connect to the hardware via an ISP that customers would be responsible for. The introduction of multiple external parties (Amazon EC2 & Company C's ISP) would result in additional complexity whilst troubleshooting as multiple external parties would be involved.

- In the 'as-is' situation, the customer is dependent upon Company B as it has exclusive control of the hardware on which the IT solution runs.

- In the 'to-be' situation, Company B loses an element of customer lock-in as does not have direct control over the hardware that runs customers' IT solutions thereby opening the possibility of a competitor bidding for support work that Company B currently provides at a reasonable margin of profit.

Following these observations we performed risk identification by using hazard keywords to identify and understand the risks posed by each inter-organizational dependency. We identified 14 specific risks associated with relying on a commercial cloud infrastructure service provider. These included some perhaps unexpected risks including:

- Insufficient or inaccurate cloud infrastructure documentation leading to an inability to respond to support calls in a timely manner rendering the service unmanageable on cloud infrastructure. Company B may be liable for breaches of SLA.

- The delayed creation, or inability to create, a virtual machine instance resulting in an inability to provide services to customers. Company B may be liable for breaches of contract for not delivering the solution according to agreed schedule.

- Changes to cloud pricing models (or being told of changes after the event) resulting in billing models that no longer corresponding to the actual charges resulting in potential financial loss.

- Changes in cloud service offerings resulting in service disruption, degradation of service, or increases in cost. Changes in cloud service offerings resulting in products having been sold that are no longer deliverable. Company B may be liable for breach of contract with customers.

- Competitors offering rival support services due to the openness of the cloud infrastructure.

- The customer's ISP becoming impaired resulting in Company B's engineers having to troubleshoot the problem with multiple external parties despite no wrong doing on their part.

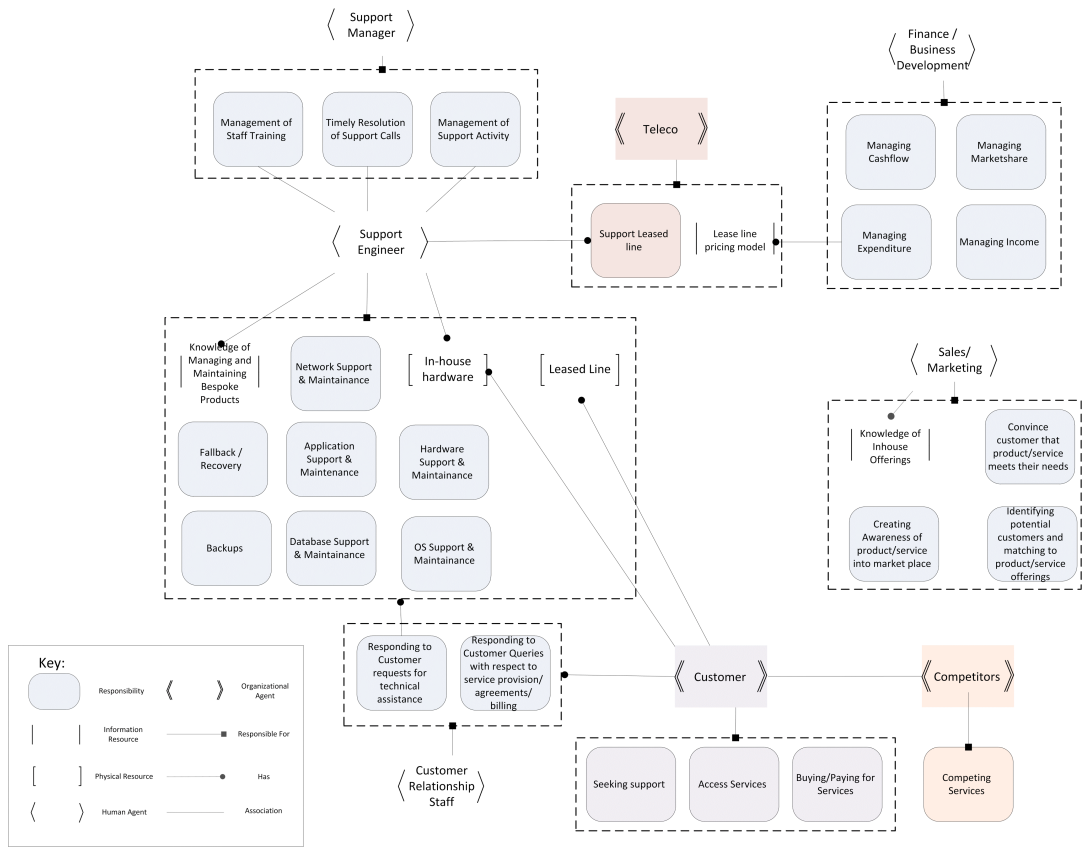

Figure 3.  Responsibility model of the "as-is" system

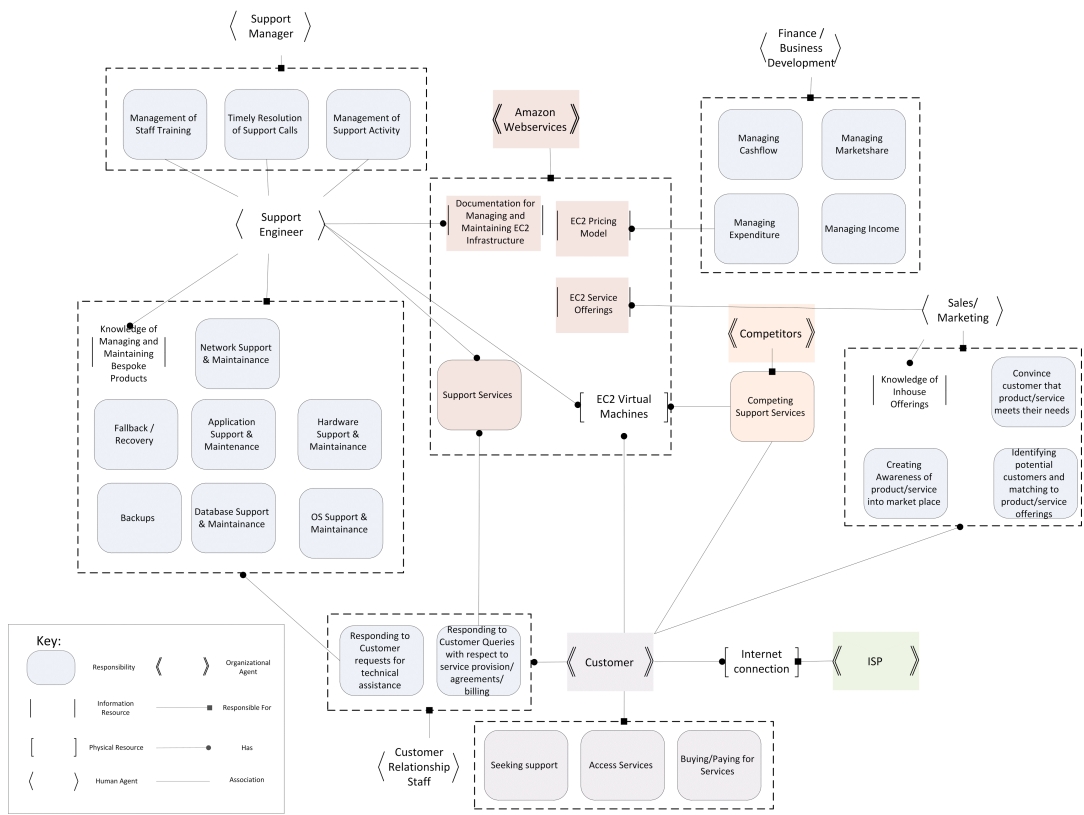

Figure 4.  Responsibility Model of the "to-be" situation

TABLE 1 EXAMPLE 'RISK CLAUSE' FROM COS RISK ANALYSIS

| Target | Hazard Keyword | Condition | Consequences / Liabilities | Risk (Li/Sev) | Recommended Action |
|---|---|---|---|---|---|
| Documentation for Managing and Maintaining EC2 | Insufficient | Documentation does not provide sufficient or adequate knowledge of EC2 infrastructure to maintain a commercial data acquisition systems | Data acquisition system is not maintainable on EC2. Timely resolution of support calls is not manageable on EC2. Liable for breach of SLA with customer. | Low/High | Assess adequacy of documentation prior to migration and perform pilots to minimize risk. Renegotiate customer support SLAs with customer |
| EC2 Service Offering | Changes | EC2 services being used to support customers are withdrawn | Customer may have service disrupted or service degradation resulting in SLA liabilities. Increase in support calls. Liable to breach of contract for services sold that are undeliverable. | Low/High | Find alternative way of provisioning service to customers. Consider implementing back-out plans to a different infrastructure. |

## V. CONCLUSIONS AND FUTURE WORK

This case study demonstrates that responsibility modeling when coupled with hazard/risk-based keywords provides a means of identifying sociotechnical risks associated with coalitions-of-systems. It exposes the risks and liabilities an organization could face if coalition partners renege on their responsibilities to provision parts of a service or application.

We are currently extending the approach to expand the scope of analysis to include the viewpoints and values of the stakeholders that will be impacted by migrating to a coalition-of-systems. This will add an additional dimension to the risk analysis, as it will go beyond looking at risks from a mechanistic means-end perspective where human agents are assumed to be passive & compliant. Instead it will also take into account factors that make stakeholders resist and conflict with change, which we believe represents an important class of risk that is missing from this analysis.

ACKNOWLEDGMENT

We would like to thank our colleagues, especially Ali Khajeh-Hosseini, at the LSCITS project (www.lscits.org) for their contributions and support.